\documentclass[12pt,preprint]{aastex}         
\def\be{\begin{equation}}
\def\ee{\end{equation}}

\def\lambar{\lambda\llap {--}}

\def\lambar{\lambda\llap {--}}

\def\lsim{\lower 2pt \hbox{$\, \buildrel {\scriptstyle <}\over
         {\scriptstyle \sim}\,$}}
\newcommand\gsim{\buildrel > \over \sim}
\begin{document}
\newcommand{\figureout}[2]{ \figcaption[#1]{#2} }       

\title{Particle Acceleration in Pair-Starved Pulsars}

\author{Alex G. Muslimov\altaffilmark{1,2} \& 
Alice K. Harding\altaffilmark{2}}   

\altaffiltext{1}{Present address: ManTech International Corporation, 
Lexington Park, MD 20653}

\altaffiltext{2}{Laboratory of High Energy Astrophysics,      
NASA/Goddard Space Flight Center, Greenbelt, MD 20771}
 

\begin{abstract}
We investigate the physical situation above the pulsar polar cap (PC) where the accelerating 
primaries (electrons) are not capable of producing sufficient numbers of electron-positron pairs 
at low altitudes (within $\sim $1-2 stellar radii above the PC surface) to screen the accelerating 
electric field, $E_{\parallel }$, and continue accelerating up to, at least, the very high 
altitudes nearly approaching the light cylinder. We derive an analytic solution for the 
$E_{\parallel }$ valid at high altitudes. The solution is based on the physical condition of 
asymptotic vanishing of the rotationally induced transverse electric field within the magnetic flux 
tube. This condition constrains the asymptotic value of the effective space charge 
that determines the distribution of electrostatic $E_{\parallel }$ within the magnetic tube. Our 
estimates of low- to high-altitude values of $E_{\parallel }$ imply the occurrence of a regime of 
primary acceleration (with the characteristic Lorentz factor up to $\sim 1-2\times 10^7$) all the 
way from the PC to the light cylinder limited by curvature-radiation reaction. In this model 
the primary outflow becomes asymptotically force-free, and may turn into a relativistic wind 
beyond the light cylinder. Such a solution will apply to both older pulsars producing only inverse 
Compton scattering pairs and younger very high $B$ pulsars (magnetars). We suggest that pulsars, 
which are lying below the pair death line, may be radio-quiet $\gamma -$ray sources.  
\end{abstract} 


\pagebreak
  
\section{INTRODUCTION}

It is believed that one of the fundamental properties of isolated pulsars is the occurrence 
of electron-positron pairs (see Sturrock 1971, Ruderman \& Sutherland 1975, and Arons \& Scharlemann 1979) 
above their polar caps (PCs).  Recent detailed investigation of the accelerating conditions in isolated 
neutron stars (NSs) implies, however, that the acceleration 
of primary particles above pulsar PCs may not necessarily result in electron-positron pair formation 
(see e.g. Harding \& Muslimov 1998,[HM98]; 2001,[HM01] and 2002,[HM02] Hibschmann \& Arons 2001, and 
Harding, Muslimov \& Zhang 2002 [HMZ02]).   HM02 studied magnetic pair production by both curvature radiation
(CR) and inverse Compton (ICS) photons and the possible screening of the accelerating electric field by pairs.
They showed that younger pulsars (with ages $\tau \lsim 10^7$ yr) are capable of producing sufficient numbers of pairs through CR to completely screen the electric field at low altitudes, while many
older pulsars are able to produce only pairs through ICS in numbers that are insufficient for screening.
For these older radio pulsars below the CR death line, as well as an unknown number of even older 
(and radio quiet) pulsars below the ICS death line,
the primaries may keep accelerating and emitting high-energy photons up to very high altitudes without any 
significant pair production.  In the present study we develop the electrodynamic model which may be applicable 
to the description of high-altitude primary acceleration and $\gamma $-ray emission in both pair-starved 
radio pulsars and pairless, radio-quiet pulsars. We derive a simple analytic solution corresponding to the 
steady-state flow of primaries and discuss the physical 
constraint on the solution near and beyond the light cylinder. This constraint determines the high-altitude 
solution that can be analytically matched with the known solution at low altitudes. 
This acceleration model can be used for older ($\tau \gsim 10^7$ yr) radio pulsars, including the majority of
millisecond pulsars, as well as pulsars that 
are lying below the theoretical pair death lines (say in the ${\dot P}$-$P$ diagram) and may manifest themselves 
only as $\gamma $-ray sources (unidentified EGRET $\gamma $-ray sources?). 
Thus, the mode of operation of an isolated NS with dominated by high-energy emission and no pairs can be tested observationally.   

Some of the unidentified, and so far, radio quiet, $\gamma$-ray sources detected by the EGRET telescope of CGRO 
have been associated with the Gould Belt (Grenier 2000, Gehrels et al. 2000). 
These faint sources have softer spectra and seem to be different from 
unidentified EGRET sources along the Galactic plane and from the known $\gamma $-ray pulsars. 
Harding \& Zhang (2001,[HZ01]) explored the possibility that radio-loud $\gamma $-ray pulsars viewed at a large 
angle to the NS magnetic pole could contribute to this new population of Galactic unidentified EGRET sources.  
They showed that the bulk of the Gould-Belt $\gamma$-ray sources could be pulsars if their emission is 
seen at large angles to their magnetic axes, such that we are missing the radio emission and hard $\gamma$-ray 
emission at lower altitudes but detecting only the higher-altitude off-beam emission. 
Due to the flaring of the dipole field lines, this emission may be seen over a large 
solid angle, far exceeding that of the main beams.  HZ01 and Harding et al. (2004) estimated that many
of the EGRET sources associated with the Gould Belt may be such off-beam $\gamma$-ray pulsars.
The population of intrinsically radio-quiet pulsars below
the ICS pair death line could also be contributing to the unidentified EGRET sources in the Gould Belt.  
Their $\gamma$-ray emission, while less luminous, would also be seen over a wide range of angles and the
acceleration model developed in this paper can be used to elucidate their predicted characteristics.
We should point out (HZ01) that off-beam emission is not expected in outer 
gap models (see e.g. Yadigaroglu \& Romani 1995, Cheng \& Zhang 1998), which predict that most low Galactic latitude 
unidentified EGRET sources in the plane are radio-quiet, Geminga-like pulsars. 

An acceleration model for the younger pulsars above the CR pair death line has recently been investigated by Muslimov \& Harding (2003, 2004, [MH04]).  Even these pulsars that have large pair multiplicity and complete screening
of the electric field still have a narrow slot gap along the last open field lines where no pairs are produced 
and where the field is screened.  High altitude particle acceleration is also possible for such pulsars, but MH04 showed
that the solution for the slot gap field differs significantly from the solution presented in this paper for
high-altitude acceleration over the entire open field region above the PC. 

The paper is organized as follows. In \S 2 we introduce the electrodynamics of relativistic primaries 
accelerating above the polar cap of the NS. The constraint on high-altitude effective space 
charge is discussed in \S 3. In \S 4 we present the solution for the accelerating electric field valid at 
high altitudes and discuss the construction of general solution extending from the polar cap surface up to 
nearly the light cylinder. Finally, in \S 5 we discuss implication of our result for $\gamma $-ray pulsars, 
and draw our principal conclusions.

\section{The Main Electrodynamic Equations}

Let us consider the magnetosphere of a rotating NS and assume that in the frame of reference rigidly corotating 
with the NS the magnetic field is stationary. Then, the general relativistic Maxwell's equations yield 
(see MT92, MH97)
\be
{\bf E} + {1\over \alpha c} ({\bf u}-{\bf w})\times {\bf B} = 
- {1\over \alpha }{\bf \nabla } \Phi ,
\label{gradPhi}
\ee
where {\bf E} and {\bf B} are the electric and magnetic fields defined in 
Zero-Angular-Momentum-Observer (ZAMO) frame of reference (see Macdonald \& Thorne 1982 for more detail); 
${\bf u} = {\bf \Omega} \times {\bf r}$ and {\bf w} are the velocity of the frame of reference rigidly corotating with a NS
($\Omega $ is the stellar angular velocity), and the differential velocity of rotation of inertial frame of 
reference, respectively, relative to the distant observer; $\Phi $ is the scalar potential; $\alpha = \sqrt{1-\varepsilon /\eta }$ ($\varepsilon = r_{\rm g}/R$, $r_{\rm g}$ is the gravitational radius of NS of radius R, and $\eta = r/R$) is the so-called general-relativistic `lapse function';  
and operator $\nabla $ should be taken in curvilinear coordinates. Taking the divergence of eq. (\ref{gradPhi}) and making use of Maxwell equation 
\be
{\bf \nabla } \cdot {\bf E} = 4 \pi \rho ,
\label{divE}
\ee 
we get the Poisson's equation for the scalar potential $\Phi $
\be
{\bf \nabla }\cdot \left( {1\over \alpha} \nabla \Phi \right) = - 4 \pi (\rho - \rho _{_{ \rm GJ}} ),
\label{Poisson}
\ee
where 
\be
\rho _{_{\rm GJ}} = - {1\over {4 \pi c}} {\bf \nabla }
\cdot \left[ {1\over \alpha } ({\bf u}-{\bf w}) \times 
{\bf B} \right] 
\label{rhoGJ}
\ee
is the general relativistic counterpart of the Goldreich-Julian (GJ) charge density 
(cf. Goldreich \& Julian, 1969), and $\rho $ is the actual charge density of electrons determined by their relativistic flow along the magnetic field lines and which is fixed by the condition $E_{\parallel } = 0$ at the stellar surface (see MT92 for details). 

In this paper, in most of our calculations (unless otherwise stated) we assume that the NS's magnetosphere has 
two distinctive regions: the corotating region with closed field lines 
filled with charged plasma, and the open-field line region extending beyond the maximum radius of corotation 
(in cylindrical coordinates with the $z$-axis along the NS's rotation axis), the light-cylinder radius 
$R_{\rm lc}\approx c/\Omega $. As in our previous studies, we 
will be working in magnetic spherical polar 
coordinates ($\eta = r/R, \theta , \phi _{pc}$), and denote by $\chi$ the pulsar obliquity (angle between the 
NS rotation axis and magnetic dipole moment). We will also refer to a `normal polarity' pulsar as one having 
$0^{\circ} \leq \chi < 90^{\circ}$ (north magnetic pole near north astrographic pole: 
${\bf \Omega} \cdot {\bf m} > 0$, where $\bf m $ is the NS magnetic dipole moment), and a `reversed polarity' 
pulsar as one having $90^{\circ} < \chi \leq 180^{\circ}$ (north magnetic pole near south astrographic pole: 
${\bf \Omega }\cdot {\bf m } < 0$). 

In coordinates ($\eta, \theta , \phi _{pc}$) eq.(\ref{Poisson}) takes the following form 
\be
{1 \over {R^2\eta ^2}}\left[ \alpha {\partial \over {\partial \eta }}\left( \eta ^2 {{\partial \Phi }
\over {\partial \eta }} \right) + {1\over \alpha } \nabla _{\Omega }^2 \Phi \right] = 
- 4 \pi (\rho - \rho _{_{\rm GJ}}),
\label{Poisson-expl}
\ee
where
\be
\nabla _{\Omega }^2 = {1\over {\sin \theta }} {{\partial }\over {\partial \theta }}\left( \sin \theta 
{\partial \over {\partial \theta }} \right) + {1\over {\sin ^2 \theta }} {{\partial ^2}\over 
{\partial \phi _{\rm pc}^2}}
\label{ang-laplace}
\ee
is the angular part of the Laplacian.

Note that the LHS of eq. (\ref{gradPhi}) can be treated as the electric field seen by the observer who is rigidly 
corotating with the NS (who moves with the velocity $\bf u - w$, i.e. with the magnetic field lines within the 
light cylinder),   
\be
{\bf E'} = {\bf E} + {1\over {\alpha c}} ( {\bf u} - {\bf w} ) \times {\bf B} .
\label{E'}
\ee
In our model calculation we shall assume that the magnetic field has pure dipole geometry, at least well within the 
light cylinder. It is important that the static-dipole magnetic field and the small-angle approximation (see below) become 
inaccurate for a rotating NS magnetosphere at very high altitudes as we approach the light cylinder. 
In addition, as we approach the light cylinder, we will see (in ZAMO frame of reference) a strong transverse electric field induced by the rotation of the magnetic flux tube. Near and beyond the light cylinder the magnetic flux tube should rather rotate at a retarded $u < c$, sweeping back in the direction opposite to rotation, thus partially reducing the electric field induced by rotation. Also, this electric field may be reduced by the effect of particle drift across the magnetic flux 
tube, but this effect is less pronounced for the entire open field line region compared to e.g. the narrow slot-gap region (MH04), where this drift may dominate. Note that, at the very light cylinder, where the velocity of the magnetic tube is formally approaching, and becomes equal to, the velocity of light, the effective transverse electric field as seen by 
the observer corotating within the magnetic tube ${\bf E}_{\perp }+ c^{-1}({\bf u} - {\bf w})\times {\bf B}$ should vanish to ensure that ${\bf E'}$ is finite, because ${\bf E'}_{\perp } \rightarrow \Gamma [{\bf E}_{\perp }+ c^{-1}({\bf u} - {\bf w})\times {\bf B}]$, where $\Gamma \approx [1-(u/c)^2]^{-1/2} \rightarrow \infty $ as $u \rightarrow c$. More generally, we suggest that the condition of vanishing of 
${\bf E}_{\perp }+ c^{-1}({\bf u} - {\bf w})\times {\bf B}$ is the basic physical condition (see also MH04) for sustaining the continuous steady-state flow within the magnetic flux tube and which also ensures that the last open field line surface will remain equipotential (MH04). Note that $\rho $ (as well as all other electrodynamic quantities) is defined in ZAMO frame of reference, and at higher altitudes 
it is mostly determined by the rotation of the magnetic flux tube. Asymptotically, at higher altitudes, the ZAMO sees the electric field 
dominated by the (transverse) rotation-induced component ${\bf E} \approx -({\bf u}/c)\times {\bf B}$, which is 
associated with nearly GJ (fictitious) charge density, $\rho = (1/4\pi ) \nabla \cdot {\bf E} \approx - (1/4\pi c)\nabla \cdot ({\bf u}\times {\bf B}) = \rho _{_{\rm GJ}}$. Thus, in a steady state, within the magnetic flux tube, the transverse electric field induced by rotation of the tube vanishes, 
\be
{\bf E}_{\perp } + {{\bf u}\over c} \times {\bf B} \approx 0,
\label{Eperp}
\ee
and the electric field within the magnetic tube is essentially determined by the distribution of the actual space charge within the tube. Here, for brevity, we have assumed that $u \gg w$ at high altitudes. The condition (\ref{Eperp}) allows us to derive the high-altitude solution for $E_{\parallel }$ without explicit calculation of $\rho $ (and also $B$) at very high altitudes. In contrast, at very low altitudes the rotation only slightly perturbs the steady-state solution and may result in establishing a weak current along the boundary field lines without violating the equipotentiality of the boundary (see discussion followed by eq. [35] in MH04). In this case, $\rho $ is simply determined by $\rho _{_{\rm GJ}}$ at the stellar surface, and the electric field, $\bf E$, is mostly determined by the distribution 
of physical charges ejected from the PC surface. Note that condition (\ref{Eperp}) is called the `ideal MHD condition' 
(see e.g. Mestel, 1999), though here it simply reflects the induction of an electric field by a rotating magnet and may be valid even for a single test particle moving along the field 
line. [Strictly speaking, the MHD approximation is only valid in collisional plasmas and also may fairly 
well describe the perpendicular (but not the parallel!) motions of collisionless plasmas.]   
    
Above the PC surface (and up to very high altitudes) the radiation drag (e.g. due to 
curvature radiation) may become important, and the motion of primaries along the magnetic field lines can ignore  
inertia and be satisfactorily approximated by the balance of $eE_{\parallel }$ and the drag force. In the end of 
Section 4 we use this approximation (see also MH04) to estimate the characteristic Lorentz factor of a primary 
moving in the regime limited by the curvature-radiation reaction.  
  
\section{The Constraint on Effective Space Charge at High Altitudes}

The condition (\ref{Eperp}) implies that at very high altitudes, near the light cylinder, maintaining a 
steady-state and continuous flow of charged particles along the magnetic stream lines requires that  
the effective space charge density, $|\Delta \rho | = |\rho _{_{\rm GJ}} - \rho |$, remains $\ll |\rho _{_{\rm GJ}}|$. 
We assume therefore that the electrostatic solution which will be valid up to high altitudes within the light cylinder should comply with the asymptotic condition (\ref{Eperp}). In this Section we shall illustrate that the 
high-altitude expression for the effective space charge density can be determined by employing the standard 
expression for $\rho $ valid within relatively low altitudes (or, for the static magnetic field) and  
$\rho _{_{\rm GJ}}$ calculated for pure dipole magnetic field rigidly co-rotating with the star. The only 
constraint we need is the asymptotic vanishing of ${\bf E}_{\perp } +{\bf u}\times {\bf B}/c$ along 
the magnetic stream lines. It proves instructive to use the effective flux of charges (both actual and GJ) 
through the cross-sectional area of a magnetic flux tube (of dipole field) emanating from the PC, defined as (see MH04) 
\be
{\cal S}(\eta ) = {\cal S}(1) {{f(1)}\over {f(\eta )}} \eta ^3 ,
\label{S}  
\ee
where $f(\eta )$ is the general-relativistic correction factor (see Appendix A). 

The flux of charges streaming with relativistic velocity through ${\cal S}(\eta )$ can be written as  
\be
F = \alpha (\eta )c {\cal S}(\eta ) \rho (\eta ),
\label{F}
\ee 
where $\rho $ is the local charge density (as seen by ZAMO). Then, using the explicit expressions for $\rho $ 
(valid for low altitudes/slow rotation and coinciding with the actual charge density seen in the frame of reference 
corotating with the magnetic flux tube) and $\rho _{_{\rm GJ}}$, for the flux of actual and GJ charges we get the approximate (for illustration) expressions (see e.g. eqs [20], [21] in MH04)
\be
F^{\rm d} = F_0 \left[ \left( 1-\kappa - {3\over 2} {{\xi ^2 }\over {\eta _{\rm lc}}} \right) 
\cos \chi + {3\over 2} {{\xi }\over {\sqrt{\eta _{\rm lc}}}} \sin \chi \cos \phi _{\rm pc} 
\right]
\label{Fd}
\ee  
and 
\be
F^{\rm d}_{_{\rm GJ}} = F_0 \left[ \left( 1-{\kappa \over {\eta ^3}} - {3\over 2} {\eta \over {\eta _{\rm lc}}}
\xi ^2 \right) \cos \chi + {3\over 2} \sqrt{{\eta \over {\eta _{\rm lc}}}}\xi \sin \chi \cos \phi _{\rm pc}, 
\right]
\label{FdGJ}
\ee
respectively. Here the superscript `d' refers to the dipole field;  
$F_0 = - \rho _0 S(\eta = 1) c$, $\rho _0 \equiv \Omega B_0/2\pi c$ and $\kappa $ is the magnitude 
of the general-relativistic effect of frame dragging at the stellar surface measured in stellar angular velocity 
$\Omega $. For most more or less realistic NS equations of state, $\kappa \approx 0.15~I_{45}/R_6^3$; where 
$I_{45}=I/10^{45}$ g$\cdot $cm$^2$, $R_6=R/10^6$ cm, $I$ is the moment of inertia of NS of radius $R$). 

\noindent 
In expressions (\ref{Fd}) and (\ref{FdGJ}) we omitted the general-relativistic red-shift corrections (factors 
$H$ and $f$ which are presented in Appendix A) and factor $\sqrt{1-3\xi ^2\eta /4\eta _{\rm lc}}$ in the expression 
for $\rho $ that takes into account the weakening of the dipole magnetic field strength towards the magnetic equator 
(see eq. [21] in MH04).   

At the very high altitudes, nearly approaching the light cylinder, the magnetic stream lines begin 
deviating from those of a vacuum magnetic dipole, and, more importantly, the terms associated with the 
rotation become the leading terms in the expression for $F$ (which will be different from formula [\ref{Fd}] based on the static magnetic 
dipole approximation), so that $F\rightarrow F_{_{\rm GJ}}$ to comply with condition (\ref{Eperp}). We shall now 
demonstrate this on a simple model. We assume that near the light cylinder the open magnetic field lines can be roughly 
approximated as radial lines. This assumption seems to be reasonable and may correspond e.g. to the 
situation where beyond the light cylinder the kinetic energy density of particles (primary electrons) 
exceeds the magnetic energy density, so that the magnetic field lines become 
governed by the outflowing particles. Or, there may be a regime of force-free current discussed by Contopoulos, 
Kazanas, \& Fendt (1999), and Mestel (1999) who neglected the particle inertia and assumed that the local current 
density is nearly equal to the GJ current density. As a result, in both cases, the magnetic field lines (in open 
field lines domain) of an initially dipole magnetic field are expected to straighten out near (or even well within) 
the light cylinder to become quasi-radial and then become dominated by the azimuthal component further out beyond the light 
cylinder. For the radial magnetic field, ${\bf B} = B_{\rm r} {\hat {\bf e}}_{\rm r}$ (in spherical coordinates 
$r$, $\theta $, $\phi _{\rm pc}$, and with unit vector in radial direction ${\hat {\bf e}}_{\rm r}$) we can get the following expression for the flux of GJ charges (here, for the sake of illustration, we ignore correction factors 
attributed to the static general-relativistic effects)
\be
F_{_{\rm GJ}}^{\rm rad} = F_0 (\cos \chi \cos \theta + \sin \chi \sin \theta \cos \phi _{\rm pc}).
\label{FrGJ}
\ee
In derivation of formula (\ref{FrGJ}) we assumed that the open field lines of the poloidal magnetic field become 
nearly radial at $\eta \gsim \eta _{\rm B}$, and that 
\be
B_{\rm r} = {B_0 \over {\eta  _{\rm B}^3}} \left( {{\eta _{\rm B}}\over \eta } \right) ^2, 
\label{Br}
\ee 
where $B_0$ is the surface strength of dipole magnetic field at the magnetic pole. 
Note that eq. (\ref{FrGJ}) implies that the magnetic stream lines are radial lines ($\theta = {\rm const}$), 
whereas eq. (\ref{FdGJ}) implies that, on magnetic stream lines (see formula [\ref{sin-theta}] below),  
$\theta = \sin ^{-1} (\xi \sqrt{\eta/\eta _{\rm lc}})$. From expressions (\ref{FdGJ}) and (\ref{FrGJ}) one can see that $F_{_{\rm GJ}}^{\rm d}$ can satisfactorily 
match the $F_{_{\rm GJ}}^{\rm rad}$ for $\eta \lsim (1/3) \eta _{\rm lc}$ and $\theta \lsim 1/3$. Due to the 
charge continuity equation, the quantity $\rho /B$ (for static magnetic field) should be constant along the magnetic 
stream lines. This means that for the static magnetic field with dominating radial component, 
$\rho \propto B \propto \eta ^{-2}$, and therefore $F^{\rm rad}$ ($\propto \rho \eta ^2$) should also 
be constant along the magnetic stream line. Even though this result is trivial, it illustrates the general fact that, in a steady state, 
the flux of charges streaming within the magnetic tube remains constant and does not depend on the geometry 
of the tube. To derive a general form of $F$ at all altitudes which complies with condition (\ref{Eperp}), we 
match $F^{\rm d}$ with $F_{_{\rm GJ}}^{\rm rad}$ at high altitudes. In fact, this very general reasoning 
allows us to restore the high-altitude flux $F$ seen by ZAMO. For example, 
for $F$ we can write the following expression containing the same radial dependence as in 
$F_{_{\rm GJ}}$ at $\eta \gsim \eta _{\rm c}$ (where $\eta _{\rm c }$ is some adjustment radial coordinate 
to be determined),
\begin{eqnarray}
F & = & F_0 \left\{ \left[ 1-\kappa -\kappa \left( {1\over {\eta ^3}} - {1\over {\eta _{\rm c}^3}}
\right) h(\eta - \eta _{\rm c}) -{3\over {2\eta _{\rm lc}}}[ (\eta - \eta _{\rm c})h(\eta -\eta _{\rm c}) + 1]
\xi ^2 \right] \cos \chi + \right. \nonumber \\
& & \left. {3\over 2} \left[ \left( \sqrt{{\eta \over {\eta _{\rm lc}}}} - \sqrt{ {{\eta _{\rm c }}\over 
{\eta _{\rm lc}} } } \right)h(\eta - \eta _{\rm c })\xi + {\xi \over {\sqrt{\eta _{\rm lc} } } } \right]  
\sin \chi \cos \phi _{\rm pc}\right\}, 
\label{Fd2} 
\end{eqnarray}
where, for simplicity, we used step function $h(x)$ ($h(x) = 1$ if $x \geq 0$; $0$, otherwise), so that 
for $\eta < \eta _{\rm c }$ we arrive at formula (\ref{Fd}), while for $\eta \geq \eta _{\rm c }$ we get 
the asymptotic expression for $F$. Note that for our purpose the specific functional form of adjustment 
of $F^{\rm d}$ to its asymptotic expression at very high altitudes is not important, because the only physical 
quantity we actually need is the effective flux, $F_{_{\rm GJ}}^{\rm d}-F$,  within the magnetic tube. Since 
the effective flux is constant at higher altitudes, it can be determined just at very high altitudes where 
asymptotic condition (\ref{Eperp}) is satisfied. Thus, using expressions (\ref{FdGJ}) 
and (\ref{Fd2}) we can get 
\begin{eqnarray}
\Delta F & \equiv & F_{_{\rm GJ}}^{\rm d} - F = F_0 \left\{ \left[ \kappa \left( 
1-{1\over {\eta ^3}} - \left( {1\over {\eta _{\rm c }^3}} - {1\over {\eta ^3}} \right) h(\eta -\eta _{\rm c }) 
\right) - \right. \right. \nonumber \\
& & \left. \left. {3\over {2\eta _{\rm lc}}} [\eta -1 - (\eta - \eta _{\rm c})h(\eta - \eta _{\rm c})] \xi ^2  
\right] \cos \chi + \right. \nonumber \\
& & \left. {3\over {2\sqrt{\eta _{\rm lc}}}} \left[ \sqrt{\eta } -1 - (\sqrt{\eta } - \sqrt{\eta _{\rm c }}) 
h(\eta - \eta _{\rm c}) \right] \xi \sin \chi \cos \phi _{\rm pc} \right\}.
\label{DF}
\end{eqnarray}  
For $\eta < \eta _{\rm c}$ formula (\ref{DF}) reduces to 
\begin{eqnarray}
\Delta F & = & F_0 \left\{ \left[ \kappa \left( 1 - {1\over {\eta ^3}} \right) + 
{3\over 2} \theta _0^2 (1- \eta ) \xi ^2 \right] \cos \chi \right. \nonumber \\
& & \left. {3\over 2} \theta _0 \xi (\sqrt{\eta } -1) \sin \chi \cos \phi _{\rm pc} \right\}, 
\label{DF1}
\end{eqnarray} 
and for $\eta \geq \eta _{\rm c}$ we get 
\begin{eqnarray}
\Delta F & = & F_0 \left\{ \left[ \kappa \left( 1 - {1\over {\eta _{\rm c}^3}} \right) + {3\over 2} 
\theta _0^2 (1-\eta _{\rm c}) \xi ^2 \right] \cos \chi  + \right. \nonumber \\
& & \left. {3\over 2} \theta _0 ( \sqrt{\eta _{\rm c}} - 1 ) \xi \sin \chi \cos \phi _{\rm pc} \right\},
\label{DF2}
\end{eqnarray}
where the general-relativistic factors $f(1)$ and $H(1)$ are temporarily omitted for the sake of simplicity. 

\noindent
Formulae (\ref{DF1}) and (\ref{DF2}) turn out to be very general and illustrate that $\Delta F$ gradually 
varies from low- (eq. [\ref{DF1}]) to high-altitudes (eq. [\ref{DF2}]), and that 
at very high altitudes $\Delta F$ is fully determined by $F_{_{\rm GJ}}^{\rm d}-F^{\rm d}$ fixed at low altitudes 
and perfectly conforms to condition (\ref{Eperp}). Also, formula (\ref{DF2}) implies that $\Delta F$ 
should saturate with altitude (see also MH04), and that this saturation occurs well within the light cylinder. 
Using the high-altitude expression for $\Delta F$ given by (\ref{DF2}) we can now find the appropriate 
solution for $\Phi $ which will depend on the adjustment parameter, 
radial coordinate $\eta _c$. Then, by matching 
the low-altitude solution for $\Phi $ with the corresponding high-altitude solution, we may determine the 
parameter $\eta _c$ (see also MH04). We must note that for estimating e.g. the energetics of acceleration of 
primaries we need to know only the low- and high-altitude solutions, since $|\Phi |$ is a monotonically 
increasing function of $\eta $ along the magnetic stream lines.     

\section{The High-Altitude Solution for Potential $\Phi $ Above the PC}

We begin with the general formula for the static dipolar magnetic field line (which is formally valid up to arbitrary high 
altitudes within the light cylinder),    
\be
\theta = \sin ^{-1} \left( \sqrt{\eta {{f(1)}\over {f(\eta )}}} \sin (\xi \theta _0)~ \right)  
= \sin ^{-1} \left( \xi \sqrt{{\eta \over {f(\eta )\eta _{\rm lc}}}} ~\right) \approx 
\sin ^{-1} \left( \xi \sqrt{\eta \over {\eta _{\rm lc}}} ~\right) ,
\label{sin-theta}
\ee
where $\theta _0 = [ \Omega R / c f(1)]^{1/2}$, $\eta _{\rm lc} = c/\Omega R$, and $\xi $ is the 
dimensionless magnetic colatitude of open field lines ($\xi = 1$ corresponds to the last open field lines, 
and $\xi = 0$ corresponds to the magnetic axis). [In a small-angle approximation, which implies that  
$\eta \xi ^2/\eta _{\rm lc} \ll 1$, we can write $\theta \approx \xi \sqrt{\eta /\eta _{\rm lc}}$.]     

At high altitudes, but still within the light cylinder, where $\nabla _{\Omega }^2 \gg \partial ^2/\partial \eta ^2$, 
we can use expression (\ref{DF2}) and relationship $\rho _{_{\rm GJ}}-\rho =\Delta F/\alpha cS(\eta )$ to rewrite   
eq. (\ref{Poisson-expl}) as follows 
\be
\nabla _{\Omega }^2 \Phi = {{{\cal F}_0}\over \eta } ({\cal A}\cos \chi + {\cal B}\sin 
\chi \cos \phi _{\rm pc}),
\label{Poisson-2}
\ee
where ${\cal F}_0 \equiv -2 \Phi _0 / f(1)$, $\Phi _0 = (\Omega R/c)B_0~R$. 
We shall introduce new dependent variables, $y_1 = \eta \Phi _1$ and $y_2 = \eta \Phi _2$, 
and independent variable $x = \cos \theta \equiv \sqrt{1-{\it a}^2\xi^2}$ (where ${\it a}= 
\sqrt{\eta /\eta _{\rm lc}}$ and approximation $\sin (\xi \theta _0) \approx \xi \theta _0$ is used). Then, the expressions for $\cal A$ and $\cal B$ can be written in general 
form as
\be
{\cal A} = {\cal A}_1 + {\cal A}_2 x^2,
\label{A}
\ee
\be
{\cal B} = {\cal B}_1\sqrt{1-x^2} + {\cal B}_2 x\sqrt{1-x^2}.
\label{B}
\ee
We will search for the solution of eq. (\ref{Poisson-2}) along the magnetic stream lines as having the form 
\be 
\Phi (\eta , \theta, \phi _{\rm pc}) = {\cal F}_0(\Phi _1 \cos \chi + 
\Phi _2 \sin \chi \cos \phi _{\rm pc}),
\label{Phi}
\ee
Now eq. (\ref{Poisson-2}) translates into
\be
{\partial \over {\partial x}} \left[ (1-x^2) {{\partial y_1}\over {\partial x}} \right] = {\cal A}_1 + {\cal A}_2 x^2,
\label{y1-eq}
\ee
\be
{\partial \over {\partial x}} \left[ (1-x^2) {{\partial y_2}\over {\partial x}} \right] 
- {{y_2}\over {1-x^2}}= {\cal B}_1\sqrt{1-x^2} + {\cal B}_2 x \sqrt{1-x^2}.
\label{y2-eq}
\ee
The general solutions of eqs (\ref{y1-eq}), (\ref{y2-eq}) vanishing at the last open field line 
(equipotentiality of the last open field lines, $\Phi [\xi = 1] = 0$) read 
\be
y_1 = - \left( {\cal A}_1 + {1\over 3} {\cal A}_2\right) \ln \left( {{1+x}\over {1+x_0}}  \right) 
- {1\over 6} {{\eta }\over {\eta _{\rm lc}}} {\cal A}_2 (1-\xi ^2),
\label{y1-sol}
\ee
\be
y_2 = {1\over 2} \sqrt{{{\eta }\over {\eta _{\rm lc}}}} \xi \left[ {\cal B}_1 
\left( {{1+x_0}\over {1+x}}-1  \right) + {1\over 3} {\cal B}_2 \left( x_0 {{1+x_0}\over {1+x}} - x\right) 
\right],
\label{y2-sol}
\ee
where $x_0 = \sqrt{1-{\it a}^2} \equiv \sqrt{1-\eta /\eta _{\rm lc}}$. 

In a small-angle approximation ($\eta \xi ^2/\eta _{\rm lc} \ll 1$), when $x \approx 1-[\eta /2\eta _{\rm lc}]\xi ^2$, expressions 
(\ref{y1-sol}) and (\ref{y2-sol}) reduce to
\be
y_1 \approx - {1\over 4} \left( {\eta \over {\eta _{\rm lc}}} \right) ( {\cal A}_1 + {\cal A}_2) 
(1-\xi ^2),
\label{y1-sol-appr}
\ee
\be
y_2 \approx - {1\over 8} \left( {\eta \over {\eta _{\rm lc}}} \right)^{3/2} ( {\cal B}_1 + {\cal B}_2) 
\xi (1-\xi ^2).
\label{y2-sol-appr}
\ee

\noindent
The general solution for $\Phi $ valid at high altitudes within the light cylinder, reads
\begin{eqnarray}
\Phi & = & \Phi _0 \theta _0^2 \left\{ \left[ 2 \left( 
{\cal A}_1 + {1\over 3}{\cal A}_2 \right) \left( {{\eta _{\rm lc}}\over \eta }\right) 
\ln \left( {{1+x}\over {1+x_0}} \right) + \right. \right. \nonumber \\
& & \left. \left. {1\over 3} {\cal A}_2 (1-\xi ^2)\right] \cos \chi + \left( {{\eta _{\rm lc}}\over 
\eta }\right) ^{1/2} \xi \times \right. \nonumber \\ 
& & \left. \left[ {\cal B}_1 \left( 1 - {{1+x_0}\over {1+x}} \right) +{1\over 3} {\cal B}_2 
\left( x - x_0 {{1+x_0}\over {1+x}} \right) \right] \sin \chi \cos \phi _{\rm pc} \right\}.
\label{Phi-gen}
\end{eqnarray}
In a small-angle approximation  the solution (\ref{Phi-gen}) reduces to 
\begin{eqnarray}
\Phi & = & {1\over 2} \Phi _0 \theta _0^2 \left[ ({\cal A}_1 + {\cal A}_2) \cos \chi + \right. \nonumber \\
& & \left. {1\over 2} \theta _0 \sqrt{\eta f(1)/f(\eta )} ({\cal B}_1 + {\cal B}_2) 
\xi \sin \chi \cos \phi _{\rm pc} \right] (1-\xi ^2).
\label{Phi-appr}
\end{eqnarray}

\noindent
Using expression (\ref{DF2}) we can now identify the explicit values of ${\cal A}_1$, ${\cal A}_2$, ${\cal B}_1$, 
and ${\cal B}_2$ conforming to the condition (\ref{Eperp}) and that determine the high-altitude solution for 
$\Phi $ (see also MH04), 
\be
{\cal A}_1 = \kappa \left( 1-{1\over {\eta _{\rm c}^3}} \right) + {3\over 2} (1-\eta _{\rm c}) 
{{H(1)}\over f(1)}{1\over \eta },~~~~~{\cal A}_2 = - {3\over 2} (1-\eta _{\rm c}) {{H(1)}\over f(1)}{1\over \eta },
\label{A-appr}
\ee
\be
{\cal B}_1 \approx {3\over 2} (\sqrt{\eta _{\rm c}} - 1) {{H(1)} \over {\sqrt{f(1)}}} {1\over {\sqrt{\eta }}} ,~~~~~{\cal B}_2 \approx 0.
\label{B-appr}
\ee
Here we restored general-relativistic factors $f(1)$ and $H(1)$ that were omitted in all corresponding 
expressions in previous Section used for illustration. In Appendix B we present general and approximate 
expressions for the functions determining the radial dependence of $\Phi $ and $E_{\parallel }$.  

In a small-angle approximation (this is good for small $\chi $, $\xi $ and at high altitude) we arrive at 
(cf. eq. [\ref{Phi-appr}]) 
\begin{eqnarray}
\Phi & \approx & {1\over 2} \left( {{\Omega R}\over c} \right) ^2 {{B_0}\over {f(1)}}R \left[ \kappa 
\left( 1 - {1\over {\eta _{\rm c}^3}} \right) \cos \chi + \right. \nonumber \\ 
& & \left. {3\over 4} \left( {{\Omega R}\over c} \right) ^{1/2} \lambda (\sqrt{\eta _{\rm c}}-1) \xi \sin \chi \cos \phi _{\rm pc} \right] 
(1-\xi ^2).
\label{Phi-appr-2}
\end{eqnarray}
The corresponding expression for $E_{\parallel } = - (1/R)\partial \Phi /\partial \eta $ reads
\begin{eqnarray} 
E_{\parallel } & = & E_{\parallel ,{\rm high}} \approx - {3\over 16} \left( {{\Omega R}\over c} \right) ^3 {B_0 \over f(1)} \left[ 
\kappa \left( 1 - {1\over {\eta _{\rm c}^3}} \right) (1+\xi ^2) \cos \chi + \right. \nonumber \\
& & \left. {1\over 2}(\sqrt{\eta _{\rm c}}-1) \left( {{\Omega R}\over c} \right) ^{1/2} \lambda (1+2\xi^2)\xi \sin \chi \cos \phi _{\rm pc} 
\right] (1-\xi^2).
\label{Epar-appr}
\end{eqnarray}
In expressions (\ref{Phi-appr-2}) and (\ref{Epar-appr}) $\lambda = H(1)/\sqrt{f(1)} 
\approx 1-0.63\varepsilon - 0.16\varepsilon ^2 - 0.5 \kappa (1-0.63\varepsilon - 0.21\varepsilon ^2)$. By comparing the  low-altitude solution (see eq. [\ref{E-low}] below and e.g. eq. [13] in HM98) and approximate high-altitude solution (\ref{Phi-appr-2}) one can see that the potential $\Phi $ practically saturates at $\eta \geq \eta _{\rm c}$, and the small growth of $\Phi $ occurs over a length scale of order of the light cylinder radius and is determined by terms $\propto \eta /\eta _{\rm lc}$. 

\noindent 
Note that the radial dependence in formula (\ref{Phi-gen}) is mostly determined by the effect of flaring of the 
magnetic field lines (for large $\xi $), so that near the very light cylinder (on the last open field lines), where this effect is 
especially profound our analytic solution becomes less accurate. However, for most pulsar parameters 
our solution is quite satisfactory within $\sim 0.5-0.7$ of the light cylinder radius, where approximate formulae 
(\ref{Phi-appr-2}) and (\ref{Epar-appr}) can be used.    

We can now calculate electron acceleration up to very high altitudes by employing the following simple  
formula for $E_{\parallel }$  
\be
E_{\parallel } \approx E_{\parallel , {\rm low}} \exp [-(\eta -1)/(\eta _c -1)] + E_{\parallel , {\rm high}},
\label{E-gen}
\ee
where $E_{\parallel ,{\rm high}}$ is determined by expression (\ref{Epar-appr}), and $E_{\parallel ,{\rm low}}$ 
is given by (see e.g. formula [13] in HM98) 
\begin{eqnarray}
E_{\parallel ,{\rm low}} & = & -{3\over 2} \left( {{\Omega R}\over c} \right)^2 {{B_0}\over {f(1)}}
\left[ {\kappa \over {\eta ^4}} \cos \chi + \right. \nonumber \\
& & \left. {1\over 8} \left( {{\Omega R}\over c} \right) ^{1/2} {{\beta (\eta ) }\over \sqrt{\eta }} 
\xi \sin \chi \cos \phi _{\rm pc} \right] (1-\xi ^2), 
\label{E-low}
\end{eqnarray}
where $\beta \approx 1+0.63x+0.46x^2-1.44x^3+(3\kappa /\eta ^{3.5})(1-0.83x-0.34x^2+0.61x^3)$, and 
$x = \varepsilon /\eta $. 

In formula (\ref{E-gen}) the parameter $\eta _{\rm c}$ can be determined as the best-fit parameter that 
provides a smooth transition between the low- and high-altitude values of $E_{\parallel }$. As we have 
discussed in our previous study (see MH04), construction of a general solution for $E_{\parallel }$ differs 
markedly for the regions with the ``favorably" and ``unfavorably" curved field lines introduced by 
Arons \& Scharlemann (1979). The region with favorably curved field lines corresponds 
to that with $\cos \phi _{\rm pc} > 0$ (electron acceleration for normal polarity pulsars, and positron 
acceleration for reversed polarity pulsars), and the region with unfavorably curved field lines corresponds 
to that with $\cos \phi _{\rm pc} < 0$ (positron acceleration for normal and electron acceleration for reversed 
polarity pulsars, respectively). In the classical electrodynamic model of Arons \& Scharlemann (1979) the 
accelerating electric field above the pulsar PC is determined by the component 
$\propto \sin \chi \cos \phi _{\rm pc}$, and therefore the electrons or positive charges (positrons) may 
accelerate in their model only along favorably or unfavorably curved field lines, respectively. In our 
model the presence of the component $\propto \cos \chi $ (determined by the general-relativistic 
frame-dragging effect) tends to make all field lines favorably curved in the sense that the charges of the same sign 
are accelerated outward over the PC. However, for very large obliquities and/or very short spin periods 
the field lines may become unfavorable for electron acceleration, because of the 
negative contribution from the term $\propto \sin \chi \cos \phi _{\rm pc}$ ($\cos \phi _{\rm pc} < 0$). 
Earlier (MH04) we have mentioned that for nearly aligned ($0^{\circ } \lsim \chi \lsim 60^{\circ }$) or  
nearly anti-aligned ($120^{\circ } \lsim \chi \lsim 180^{\circ }$) cases, there will be continuous acceleration 
of electrons from the PC surface up to very high altitudes on both favorably ($\cos \phi _{\rm pc} > 0$) and 
unfavorably ($\cos \phi _{\rm pc} < 0$) curved field lines. For nearly orthogonal rotators 
($80^{\circ } \lsim \chi \lsim 100^{\circ }$) there will be continuous acceleration of electrons/positrons 
from the PC surface up to very high altitudes on favorably curved field lines for normal/reversed polarity, 
and continuous acceleration of positrons/electrons on unfavorably curved field lines for normal/reversed 
polarity. At intermediate inclination ($60^{\circ } \lsim \chi \lsim 80^{\circ }$ and 
$100^{\circ } \lsim \chi \lsim 120^{\circ }$), there will be continuous acceleration of electrons on 
favorably/unfavorably curved field lines for normal/reversed polarity. Let us now summarize the characteristic 
values of $\eta _{\rm c}$ in formula (\ref{E-gen}) for the situations where in this formula either the 
term $\propto \cos \chi $ or $\propto \sin \chi \cos \phi _{\rm pc}$ dominates. We will present our estimates 
for the value of $\xi \sim 0.5$. For the case where the term $\propto \cos \chi $ in formula (\ref{E-gen}) dominates, 
$\eta _{\rm c} \sim 3-4$ and $\eta _{\rm c} \sim 10-15$ for the millisecond pulsars and relatively long-period 
($P \sim 0.5-1$ s) pulsars, respectively. For the case where term $\propto \sin \chi \cos \phi _{\rm pc}$ 
dominates, $\eta _{\rm c} \sim 3-4$ (millisecond pulsars) and $\eta _{\rm c} \sim 15-20$ (pulsars with 
$P\sim 0.5-1$ s). Thus, for a given spin period the parameter $\eta _{\rm c}$ is, generally, a function of $\chi $ and 
$\phi _{\rm pc}$.    

\noindent
Let us now estimate the characteristic electron Lorentz factor, $\gamma $, in the regime of acceleration 
above the pulsar PC limited by the CR reaction, i.e. where the condition 
\be
e|E_{\parallel}| \approx {2\over 3}{{e^2}\over {\rho _c^2}}\gamma ^4
\label{CRR}
\ee
is satisfied. Here $\rho _c \sim (4/3)~(R/\xi )~(\eta \eta _{lc})^{1/2}$ is the characteristic curvature radius 
of the field lines ($0 < \xi \leq 1$). 

\noindent
By inserting expression (\ref{Epar-appr}) for $E_{\parallel }$ into expression (\ref{CRR}), we obtain
\be
\gamma \approx 2\times 10^7 \left( {{\eta }\over {\eta _{\rm lc}}} \right) ^{1/4} 
\left( {{B_{12}R_6^3\kappa _{0.15}}\over {P_{0.1}\xi ^2}} \right) ^{1/4} 
{\Lambda (\chi, \xi , \phi _{\rm pc})}^{1/4},
\label{gamma}
\ee
\be
\Lambda = \left| \left( 1 - {1\over {\eta _{\rm c}^3}} \right) (1+\xi ^2) \cos \chi + 
{1\over 2}(\sqrt{\eta _{\rm c}}-1) \left( {{\Omega R}\over {c}} \right) ^{1/2} {{\lambda } \over \kappa } 
(1+2\xi^2)\xi \sin \chi \cos \phi _{\rm pc} \right| (1-\xi^2),
\label{Lambda}
\ee
where $B_{12} = B_0/10^{12}$ G, $\kappa _{0.15} = \kappa /0.15$, $P_{0.1} = P/0.1$ s. For the ms (e.g. $P=3$ ms, 
$B_{12} = 3\cdot 10^{-4}$, and $R_6 = 1$) and for 1 s ($B_{12} = 1$, $R_6 = 1$) pulsars we get $\gamma \sim 10^7$. 

The characteristic energy of CR spectrum, ${\cal E}_{\rm cr}$, is given by 
\be
{\cal E}_{\rm cr}/m_ec^2 = 3 {\lambar _c} \gamma ^3 / 2~\rho _c,
\label{Ecr}
\ee
where $\lambar _c = 3.9\times 10^{-11}$ cm is the electron Compton wavelength. 
By inserting expression (\ref{gamma}) into formula (\ref{Ecr}), we get 
\be
{\cal E}_{\rm cr} \approx 400~{\rm MeV}~\left( \eta \over \eta _{\rm lc} \right) ^{1/4} 
\left( {{B_{12}R_6^3\kappa _{0.15}}\over {P_{0.1}^{7/3}\xi ^{2/3}}} \right) ^{3/4} \Lambda ^{3/4}.
\label{Ecr-2}
\ee
Thus, for the ms and $\sim 1$ s pulsars ${\cal E}_{\rm cr} \sim 600$ and $\sim 10$ MeV, respectively. Thus, 
the off-beam spectrum of CR for the ms pulsars occurs at EGRET energies, but may be at a flux too small 
to detect except in the case of very nearby sources. The on-beam CR emission from ms pulsars which is 
radiated nearer the PC (Harding, Usov \& Muslimov 2004) occurs at 1-10 GeV, since $E_{\parallel ,{\rm low}} 
\gg E_{\parallel ,{\rm high}}$.   
It is interesting to point out that near the light cylinder, where the radius of curvature of open 
field lines may substantially increase, the curvature radiation may cease. However the shape of the 
curvature-radiation spectrum and peak spectral energy will be determined by the harder photons produced 
at low altitudes. 

Let us also estimate (within the small-angle approximation) the characteristic luminosity of high-energy 
off-beam emission. For example, for $\cos \chi \approx 1$, we can write
\begin{eqnarray}
L_{_{\rm HE}} & = & f_{_{\rm HE}}~(mc^2)~c~(n_er^2)~\int _0^{2\pi } d\phi _{\rm pc} \int _0^{\theta _{\rm max}} 
\gamma \sin \theta d\theta \nonumber \\
&  & \approx f_{\gamma, 1}~(mc^2)~c~(n_e\eta ^3)~(2\pi r_{\rm pc}^2)~\int _0^1 \gamma (\xi )\xi d\xi,  
\label{LHE}
\end{eqnarray}
where $f_{\gamma, 1}$ ($< 1$) is the efficiency of conversion of electron kinetic energy into high-energy (curvature) 
emission (which may be calculated by integrating the CR power), $\theta _{\rm max}= \xi \theta _0 \sqrt{\eta }$, 
$r_{\rm pc} = \theta _0 R$ is the PC radius, and $n_e \approx \Omega B_0 (1-\kappa)/(2\pi ce\eta ^3)$  is the electron number density.

By substituting expressions (\ref{gamma}) and (\ref{Lambda}) into formula (\ref{LHE}), and using 
$\kappa _{0.15} = 1$, we get
\be
L_{_{\rm HE}} \approx 2\times 10^{33}~{\rm erg}\cdot {\rm s}^{-1}~f_{\gamma, 1} {{B_{12}^{5/4}}R_6^{15/4} 
\over {P_{0.1}^{9/4}}} \left( \eta \over \eta _{\rm lc} \right) ^{1/4}.
\label{LHE2}
\ee
Note that, according to formula (\ref{LHE2}), $L_{_{\rm HE}} \propto L_{\rm sd}^{5/8}$, where $L_{\rm sd}$ 
($\sim 5 \cdot 10^{34}~{\rm erg}\cdot {\rm s}^{-1}~B_{12}^2~R_6^6~P_{0.1}^{-4}$) is the pulsar spin-down luminosity. 
We can also estimate the efficiency of conversion of pulsar spin-down luminosity into high-energy luminosity, 
\be
f_{\gamma, 2} = {{L_{_{\rm HE}}} \over {L_{_{\rm sd}}}} = 0.07~f_{\gamma, 1}~P_{0.1}^{1/4} 
\left( {\eta \over \eta _{\rm lc}} \right) ^{1/4}~L_{{\rm sd},34}^{-3/8}.
\label{LHE/LSD}
\ee
The estimate (\ref{LHE2}) is roughly consistent with the expected luminosity of off-beam pulsars (see HZ01). 
However, for the quantitative modeling of high-energy fluxes from these hypothetical sources the above formulae, 
based on a small-angle approximation, may not be accurate and the appropriate detailed treatment will be needed. 
According to (\ref{LHE/LSD}), the efficiency of high-energy emission significantly increases for long-period pulsars (both with regular, $\sim 10^{12}$ G, surface magnetic field and magnetars) and also for the very low-field ($\lsim 10^8$ G) ms 
pulsars. Note that in estimates (\ref{gamma}), (\ref{Ecr-2}), (\ref{LHE2}) and (\ref{LHE/LSD}), according to a small-angle approximation, $\eta \ll \eta _{\rm lc}/\xi ^2$ ($\xi \neq 0$). This means that for a nearly aligned pulsar and sufficiently 
small magnetic colatitudes (say $\xi \sim 0.1-0.2$) these estimates are formally valid for very large radial 
distances, $\eta \lsim \eta _{\rm lc}$. It is interesting that for small 
$\xi $ the values of $\gamma $ and ${\cal E}_{\rm cr}$ (see [\ref{gamma}] and [\ref{Ecr-2}]) are proportional 
to $1/\xi $ which implies the substantial hardening of the CR spectrum towards the magnetic axis. 
Obviously, $\xi $ cannot be arbitrary small, with the lower limit set by condition $L_{_{\rm HE}} < L_{\rm sd}$.    

In this paper we discuss the possibility of occurrence of the regime of primary acceleration up to high altitudes 
without any significant pair formation at low altitudes. However, the pairless regime of acceleration of primaries 
may not be stable (at least, for some magnetic colatitudes) against $\gamma + X \rightarrow e^++e^-$, 
$\gamma + \gamma \rightarrow e^++e^-$, or even (for magnetars) $\gamma +B \rightarrow e^++e^-$ processes. We plan 
to address this issue and calculate the emergent high-energy spectrum for the relevant scenarios separately. We expect, therefore, that the high-altitude pair formation, if it occurs, may result in effective 
narrowing of the acceleration region and associated weakening of the accelerating electric field. It is unlikely  
that the regime of pairless acceleration can be completely shut down by high-altitude pairs, though. 

\section{Discussion and Conclusions}

We have considered the possibility of continuous acceleration of primaries (electrons) on open field lines above the 
pulsar PC up to very high altitudes, nearly approaching the light cylinder. Such a possibility may occur 
in pulsars which lie below the CR pair death lines (see e.g. HM02, and HMZ02), because the efficiency of pair formation 
by ICS may not be sufficient to screen the accelerating electric field. These results will 
apply to old (age $\gsim 10^7$ yr) and ms radio pulsars, as well as to radio-quiet pulsars that are not producing 
any pairs. Well within the light cylinder the 
magnetic field is well described by pure dipole structure, and primary flow is governed by the 
magnetic field. However, as we approach the light cylinder, the particle stream lines may drastically 
deviate from the magnetic field lines of a static dipole, and beyond the light cylinder the flow of 
relativistic particles may govern the magnetic field. As a result, the distribution of GJ and actual 
charge densities near and beyond the light cylinder may significantly differ from that for low altitudes.
By considering the magnetic flux tube of a rotating NS and taking into account the fact that at high altitudes, 
the strong transverse electric field as seen by a ZAMO and which is induced by the rotation of the magnetic flux tube, vanishes within the tube, at very high altitudes the effective charge density 
$|\Delta \rho _{\rm eff}| = |\rho _{_{\rm GJ}}-\rho | \ll |\rho _{_{\rm GJ}}|$. It is this condition which provides electrostatics within the magnetic flux tube, and allows us to extend the low-altitude solution to much 
higher altitudes nearly approaching the light cylinder. Thus, we derived explicit expressions for the 
scalar potential and accelerating electric field valid through the very high altitudes, where the often-used 
small-angle approximation (condition $\xi ^2 \eta /\eta _{_{\rm lc}} \ll 1$) becomes invalid (for not very small 
$\xi $), and the corresponding low-altitude solution is not applicable. 

We have shown that the low-altitude solution for the scalar potential $\Phi (\eta )$ makes a smooth transition to 
the high-altitude solution constructed in the present study for which the corresponding charge density of 
relativistic primaries may match the wind solution akin to that derived in the force-free case (Contopoulos, Kazanas, \& Fendt 1999), 
thus connecting the near-surface pulsar electrodynamics to a global solution for the magnetosphere and wind. The high-altitude solution presented in this 
paper allows us to estimate energetics of primaries that are not capable of producing pairs at very low altitudes above 
the PC and therefore accelerate up to the very high altitudes. For example, we have demonstrated that 
these primaries may achieve significant Lorentz factors (up to a few times $10^7$) and emit $\gamma $-rays in the 
regime of acceleration limited by the CR reaction. The characteristic length scale of high-altitude acceleration 
is roughly equal to the light cylinder radius, and the high-energy luminosity scales as $L_{\rm sd}^{5/8}$.

In conclusion, we should point out that most of the difficulty in constructing the quantitative 
model for high-altitude primary acceleration arises from the lack of the reliable knowledge about the 
magnetic field structure of rotating NS near and beyond the light cylinder. We have accentuated in this study 
that, at high altitudes, the rotation of magnetic flux tube fully determines the distribution of the electric charge 
density within the tube as seen by the ZAMO (or, in a flat space-time, by the observer in the `lab' frame of 
reference), and demonstrated the electrostatics within the magnetic flux tube which is important for the calculating the high-altitude accelerating electric field. Such a regime of steady-state acceleration of primaries is crucial for  constructing a self-consistent steady-state 
global magnetospheric model.

\acknowledgments 
We acknowledge support from the NASA Astrophysics Theory Program through the Universities 
Space Research Association.

\clearpage 

\appendix
\section{General-relativistic Corrections}

In this Appendix we present explicit expressions for the physical components of the 
dipolar magnetic field around NS (in Schwarzschild metric). Also, we present some important 
general-relativistic factors entering the main electrodynamic equations in curved 
space-time. We employ the magnetic spherical coordinates $r$, $\theta $, and $\phi _{\rm pc}$,     
and use the standard notations for the fundamental physical constants. Other notations and shorthands are:

\noindent 
$R -$ the NS radius\\
$r_{\rm g} -$ the stellar gravitational radius\\ 
$\eta = r/R -$ the dimensionless radial coordinate\\
$\varepsilon = r_g/R -$ the NS compactness parameter\\
$x  = \varepsilon /\eta $\\ 
$\alpha = \sqrt{1-x} -$ the general-relativistic `lapse function'\\
$\kappa \approx 0.15~I_{45}/R_6^3 -$ the frame-dragging effect at the NS surface measured in $\Omega $, angular 
velocity of stellar rotation\\
$I_{45} = I/10^{45}~{\rm g}\cdot {\rm cm}^2$ \\
$I -$ moment of inertia of the NS \\
$R_6 = R/10^6~{\rm cm}$

The physical $r-,~\theta -$ components of dipolar magnetic field are  
\be
B_{\rm r}^{\rm d} = B_0 {{f(\eta )} \over {f(1)}} {1\over {\eta ^3}}  \cos \theta  
\label{Brad}
\ee
and
\be
B_{\theta }^{\rm d} = {1\over 2} \alpha B_0 \left[ -2 {{f(\eta )}\over {f(1)}}+{3\over 
{f(1)(1-x)}} \right] {1\over {\eta ^3}} \sin \theta ,
\label{Btheta}
\ee
respectively, where $B_0$ is magnetic field strength at the magnetic pole. Here 
\be
f(\eta )  = - {3\over {x^3}} [ \ln ( 1 - x) + x ( 1 + x/2)].
\label{f} 
\ee
For large $\eta $ (small $x$) one can use approximate formula 
\be
f(\eta ) \approx 1+0.75x +0.6x^2.
\label{fappr} 
\ee
The correction factor $H(\eta )$ entering the general-relativistic expression for 
the Goldreich-Julian charge density reads
\be
H(\eta ) = x - \kappa (x/\varepsilon )^3 + {{[1-3x/2 + (1/2)\kappa (x/\varepsilon )^3 ]}\over 
{(1-x)f(x)}}.
\label{H}
\ee
The approximate expression (for small $x$), 
\be
H(\eta ) \approx 1-0.25x -0.16x^2-0.5\kappa (x/\varepsilon )^3(1-0.25x-0.21x^2).
\label{Happr}
\ee

\clearpage 

\section{Radial Dependence of $\Phi $ and $E_{\parallel }$}

By incorporating expressions (\ref{A-appr})-(\ref{B-appr}) into eq. (\ref{Phi-gen}) we get 
\be 
\Phi = {1\over 2} \Phi _0 \theta _0^2 ( {\cal F}_1 \cos \chi + 
{\cal F}_2 \sin \chi \cos \phi _{\rm pc} ),
\label{Phi-gen-2}
\ee
where
\be
{\cal F}_1 = 4\left( {\cal A}_1 + {1\over 3} {\cal A}_2 \right) {{\eta _{\rm lc}}\over \eta } 
\ln \left( {{1+x}\over {1+x_0}} \right) + {2\over 3} {\cal A}_2 (1-\xi ^2 ),
\label{F1}   
\ee
and
\be
{\cal F}_2 = 2 \sqrt{{{\eta _{\rm lc}}\over {\eta }}} {\cal B}_1 \left( {{x-x_0}\over {1+x}} \right) \xi ,
\label{F2}
\ee
where $x = \sqrt{1-\eta \xi ^2 /\eta _{\rm lc}}$, and $x_0 = x(\xi = 1)$. 

Here we present explicit formulae for the functions ${\cal F}_1$ and ${\cal F}_2$ defined above, and their 
derivatives with respect to $\eta $. Also, we express 
${\cal F}_1$ and ${\cal F}_2$ as a power series in $\eta /\eta _{\rm lc}$, up to the quadratic terms.  
We use the same notations as in Appendix A, along with the variables\\
$\eta _{\rm lc} = c/\Omega R -$ the dimensionless (cylindrical) radial coordinate of the light cylinder \\
$\xi = \theta (\eta = 1)/\theta _0 -$ the dimensionless magnetic colatitude (see eq. [\ref{sin-theta}]) \\

The $\eta $-derivatives of ${\cal F}_1$ and ${\cal F}_2$ determine 
\be
E_{\parallel } = - {1\over 2} {\Phi _0 \over R}\theta _0^2~\left( {{\partial {\cal F}_1}\over {\partial \eta }}
\cos \chi + {{\partial {\cal F}_2}\over {\partial \eta }} \sin \chi \cos \phi _{\rm pc} \right),
\label{E-appx-B}
\ee
and can be expressed as 
\begin{eqnarray} 
{{\partial {\cal F}_1}\over {\partial \ln \eta }} & = & 2 \left( {\cal A}_1 + {1\over 3} {\cal A}_2 \right) 
\left( {1\over {x_0(1+x_0)}} - {{\xi ^2}\over {x(1+x)}} - 2 {{\eta _{\rm lc}}\over {\eta }}
\ln {{1+x}\over {1+x_0}} \right) + \nonumber \\
& & {2\over 3} {\cal A}_2 \left( 4 {{\eta _{\rm lc}}\over {\eta }} \ln {{1+x}\over {1+x_0}} - (1-\xi ^2) \right) ,
\label{F1der-appx-B}
\end{eqnarray}
\be
{{\partial {\cal F}_2}\over {\partial \ln \eta }} =  {\cal B}_1 \sqrt{{{\eta _{\rm lc}}\over {\eta }}} 
{{\xi }\over {1+x}} \left[ 2(x_0-x) - {{\eta }\over {\eta _{\rm lc}}} \left( {{1+x_0}\over {x(1+x)}}
\xi ^2 - {1\over x_0} \right) \right] .
\label{F2der-appx-B}
\ee
For $\eta /\eta _{\rm lc} \lsim 1/3$ the expressions (\ref{F1})-(\ref{F2}) can be approximated 
by the following formulae
\begin{eqnarray}
{\cal F}_1 & \approx & \left\{ \kappa\left( 1-{1\over {\eta _{\rm c}^3}} \right)\left[ 1 + 
{3\over 8} {\eta \over {\eta _{\rm lc}}}(1+\xi ^2) + {5\over 24} \left( {\eta \over 
{\eta _{\rm lc}}} \right) ^2 (1+\xi ^2 + \xi ^4)\right] + \right. \nonumber \\
& & \left. {3\over 8} {{\Omega R}\over c} {H(1) \over f(1)} (1-\eta _{\rm c}) 
\left[ 1+\xi^2 + {5\over 9} {\eta \over {\eta _{\rm lc}}} (1+\xi ^2+\xi ^4) \right] \right\} 
(1-\xi ^2),
\label{F1appr-appx-B}
\end{eqnarray}
\begin{eqnarray}
{\cal F}_2 & \approx & {3\over 4} (\sqrt{\eta _{\rm c}}-1) {H(1)\over {\sqrt{f(1)}}} 
{1\over {\sqrt{\eta _{\rm lc}}}}\left[ 1+ {1\over 4} {\eta \over {\eta _{\rm lc}}}(1+2\xi ^2) + \right. 
\nonumber \\
& & \left. {1\over 8} \left( {\eta \over {\eta _{\rm lc}}} \right) ^2 \left( 1+{3\over 2}\xi ^2 +{5\over 2} \xi ^4 
\right) \right] \xi (1-\xi^2),
\label{F2appr-appx-B}
\end{eqnarray}
\begin{eqnarray}
{{\partial {\cal F}_1}\over {\partial \eta }} & \approx & {3\over 8} {{\Omega R}\over c} 
\left\{ \kappa \left( 1 - {1\over {\eta _{\rm c}^3}} \right) \left[ 1 + \xi ^2 + {10\over 9} 
{\eta \over {\eta _{\rm lc}}} (1+\xi ^2 +\xi ^4) \right] - \right. \nonumber \\
& & \left. {5\over 9} {{\Omega R}\over c} {H(1) \over f(1)} (\eta _{\rm c}-1)(1+\xi ^2 +\xi ^4)\right\} 
(1-\xi ^2),
\label{F1der-appr-appx-B}
\end{eqnarray}
\begin{eqnarray}
{{\partial {\cal F}_2}\over {\partial \eta }} & \approx & {3\over 16} \left( {{\Omega R}\over c} \right) ^{3/2} 
(\sqrt{\eta _{\rm c}} - 1) {H(1)\over {\sqrt{f(1)}}} \times \nonumber \\
& & \left[ 1+2\xi ^2 + {\eta \over {\eta _{\rm lc}}} \left( 1 + {3\over 2} \xi ^2 + {5\over 2} \xi ^4 \right) 
\right] \xi (1-\xi ^2).
\label{F2der-appr-appx-B}
\end{eqnarray}

\end{document}